\documentclass[structabstract]{aa}  
\usepackage{graphicx}
\usepackage{txfonts}
\usepackage{natbib}
\bibpunct{(}{)}{;}{a}{}{,} 
\usepackage{color}
\newcommand{\cmtwo}{cm$^{-2}$}  
\newcommand{\cmthree}{cm$^{-3}$}

\newcommand{\kms}{km\,s$^{-1}$}       
 
\newcommand{\ecs}{erg cm$^{-2}$ s$^{-1}$}

\newcommand{\um}{$\mu$m}                                 
\newcommand{\molh}{H$_{2}$}                              

\newcommand{\water}{H$_{2}$O}

\newcommand{\molo}{O$_{2}$}
\newcommand{\lsun}{$L_{\odot}$}                          
\newcommand{\msun}{$M_{\odot}$}

\newcommand{\gapprox}{$\stackrel {>}{_{\sim}}$}   
\newcommand{\lapprox}{$\stackrel {<}{_{\sim}}$}
\newcommand{\about}{$\sim$}                       
\newcommand{\powten}[1]{10$^{#1}$}
\newcommand{\oishort}{[O\,{\sc i}]\,63\,$\mu$m}
\newcommand{\oilong}{[O\,{\sc i}]\,145\,$\mu$m}

\newcommand{\av}{$A_{\rm V}$}                     
\newcommand{\rv}{$R_{\rm V}$} 

\newcommand{\ebv}{$E_{\rm B-V}$}

\newcommand{\ro}{$\rho \, {\rm Oph}$}
\newcommand{\roc}{$\rho \, {\rm Oph \,\, cloud}$}
\newcommand{\roa}{$\rho \, {\rm Oph \, A}$}

\newcommand{\roac}{$\rho \, {\rm Oph \, A \, core}$}
\newcommand{\roca}{$\rho \, {\rm Oph \, A \, cloud}$}

\newcommand{\amin}{$^{\prime}$}                   
\newcommand{\asec}{$^{\prime \prime}$}
\newcommand{\adeg}{$^{\circ}$}

\newcommand{\radot}[4]{\mbox{#1$^{\rm h}$#2$^{\rm m}$#3$\stackrel{\rm s}
{_{\bf\cdot}}$#4}}  

\newcommand{\decdot}[4]{\mbox{#1$^{\circ}$ #2$^{\prime}$ #3$\stackrel {\prime 
\prime}{_{\bf \cdot}}$#4}}
\newcommand{\adegdot}[2]{\mbox{#1$\stackrel {\circ}{_{\bf \cdot}}$#2}}

\newcommand{\asecdot}[2]{\mbox{#1$\stackrel {\prime \prime}{_{\bf \cdot}}$#2}}

\newcommand{\vbis}{$\upsilon^{\prime \prime}$}
\newcommand{\vprim}{$\upsilon^{\prime}$}
\begin{document}
   \title{Oxygen in dense interstellar gas\thanks{Based on observations with the CAM-CVF (Cesarsky et al. 1996) and the LWS (Clegg et al. 1996) onboard the Infrared Space Observatory, ISO (Kessler et al. 1996).}}

   \subtitle{The oxygen abundance of the star forming core \roa}

   \author{R. Liseau\inst{}
          \and
          K. Justtanont\inst{}
	 }

   \offprints{R. Liseau}

   \institute{Department of Radio and Space Science, Chalmers University of Technology, Onsala Space Observatory, SE-439 92 Onsala, Sweden, 
              \email{rene.liseau@chalmers.se, kay.justtanont@chalmers.se}
	}

   \date{Received ; accepted }


  \abstract
   {Oxygen is the third most abundant element in the universe, but its chemistry in the interstellar medium is still not well understood.}
   {In order to critically examine the entire oxygen budget, we attempt here initially to estimate the abundance of atomic oxygen, O, in the only one region, where molecular oxygen, \molo, has been detected to date.}
   {We analyse ISOCAM-CVF spectral image data toward \roa\ to derive the temperatures and column densities of \molh\ at the locations of ISO-LWS observations of two [O\,I] $^3$P$_{\!\!J}$ lines. The intensity ratios of the ($J$=1-2) 63\,\um\ to ($J$=0-1) 145\,\um\ lines largely exceed ten, attesting to the fact that these lines are optically thin. This is confirmed by radiative transfer calculations, making these lines suitable for abundance determinations. For that purpose, we calculate line strengths and compare them to the LWS observations. }
   {Excess [O\,I] emission is observed to be associated with the molecular outflow from VLA\,1623. For this region, we determine the physical parameters, $T$ and $N$(\molh), from the CAM observations and the gas
   density, $n$(\molh), is determined from the flux ratio of the \oishort\ and \oilong\ lines.  For the oxygen abundance, our analysis leads to essentially three possibilities: (1) Extended low density gas with standard ISM 
   O-abundance, (2) Compact high density gas with standard ISM O-abundance and (3) Extended high density gas with reduced oxygen abundance, ${\rm [O/H]}\sim 2 \times 10^{-5}$.}
   {As option (1) disregards valid \oilong\ data, we do not find it very compelling; we favour option (3), as lower abundances are expected as a result of chemical cloud evolution, but we are not able to dismiss option (2)
    entirely. Observations at higher angular resolution than offered by the LWS are required to decide between these possibilities.}
    

   \maketitle
%

\section{Introduction}

Oxygen is the most abundant of the astronomical metals and is, as such, of profound importance for the chemistry of the interstellar medium (ISM). Therefore, its role and its relative abundance in the various phases of the ISM should of course be known and understood. \citet[][and references therein]{quan2008} present a recent overview for our understanding of the abundance of interstellar molecular oxygen. In general, the amount of \molo\ in molecular clouds has been below detection capability of the dedicated space missions SWAS \citep[e.g.,][]{goldsmith2000,bergin2000} and Odin \citep[e.g.,][]{pagani2003,sandqvist2008}. In merely one single location, viz. the dense molecular core A in the \ro iuchi cloud, did Odin detect a weak \molo\ line at the $\sim 5 \sigma$ level \citep{larsson2007}.

The oxygen abundance in the interstellar medium has been determined to be about $3.0 \times 10^{-4}$ \citep{savagesembach1996} to $3.4 \times 10^{-4}$  \citep{oliveira2005}. This is smaller than the solar value, viz. $[{\rm O/H}]_{\odot}=4.6 \times 10^{-4}$ \citep{asplund2005,allende2008}\footnote{\citet{ayres2005} and \citet{landi2007} present evidence in favour of twice that value, i.e., $8.7 \times 10^{-4}$. In a recent evaluation, \citet{melendez2008} concluded that a value of $5 \times 10^{-4}$ would be consistent with the existing different model atmospheres. There is general agreement, however, that the ratio of the {\it carbon-to-oxygen} abundance is unaffected by the different analysis techniques and remains close to one half.}, indicating that in the interstellar medium (ISM), oxygen is depleted in the gas phase.

\citet{hollenbach2009} examine theoretically the oxygen chemistry in various interstellar cloud conditions in considerable detail, taking into account processes both in the gas phase and on grain surfaces. Of particular relevance in the context of this paper would be their results applying to weakly irradiated photon dominated regions (PDRs). Weak shocks also heat (and compress) the gas which, overcoming low-temperature reaction barriers, can potentially initiate a rather complex chemistry. \citet{gusdorf2008} present results for oxygen bearing species, including \molo, which are produced by grain sputtering in C-type shock models.

The diagnostic usefulness of \molh\ lines for the study of shocked gas has been discussed in detail by \citet[][and references therein]{neufeld2006}. It had also earlier been shown that Herbig-Haro (HH) flows, which are optical manifestations of interstellar shock waves, are strong emitters of \oishort\ line emission \citep{liseau1997}. In the present paper, we will combine observations in lines of these species of the star forming core \roa.

\begin{figure}[ht]
  \resizebox{\hsize}{!}{
  \rotatebox{00}{\includegraphics{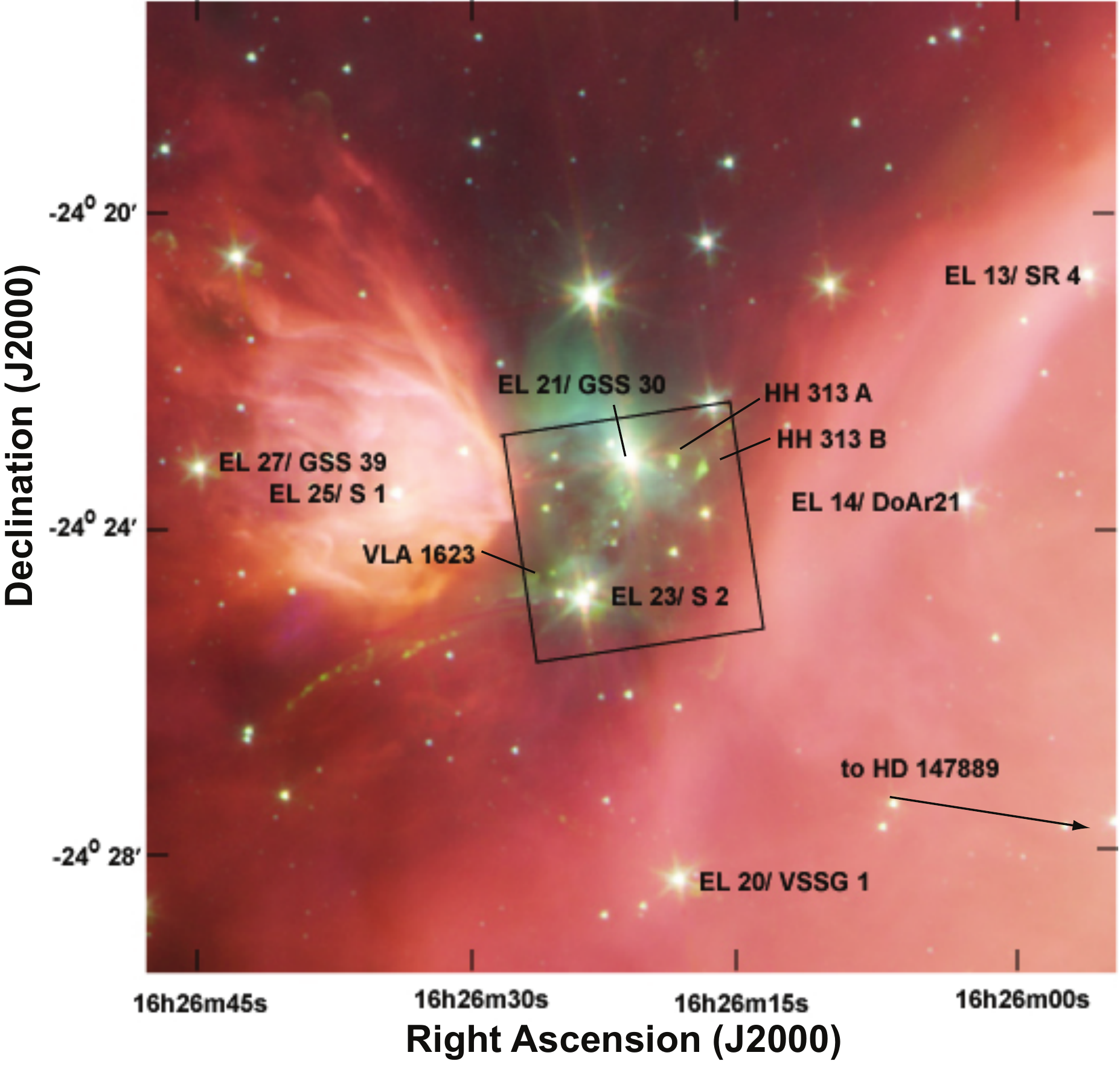}}
  }
  \caption{The orientation of the CAM-CVF frames in \roa, with the pixel coordinate (0, 0) in the upper left corner, is shown superposed onto a partial Spitzer Space Telescope image (courtesy NASA/JPL-Caltech/Harvard-Smithsonian CfA, see \texttt{http://gallery.spitzer.caltech.edu/Imagegallery/chron.}
\texttt{php?cat=Astronomical\_Images}). The colours represent effective wavelengths of 3.6\,\um\ (blue), 4.5\,\um\ (green) and 8.0\,\um\ (red). The array size of $3^{\prime} \times 3^{\prime}$ corresponds to 0.1\,pc by 0.1\,pc at the distance of the cloud \citep[120\,pc:][]{lombardi2008,snow2008}. A number of young objects are indicated, with the dominant B-star HD\,147889 being off the figure, almost 15\amin\ westsouthwest of the center of the CAM-frame. The position of the Class\,0 source VLA\,1623, which drives a collimated bipolar CO outflow \citep{andre1990}, is at the center of the dark flaring disk feature. The green colour, in particular, visualizes many flow phenomena, such as, e.g., HH\,313 and the arcs of the southeastern flow. 
  }
  \label{VLA}
\end{figure}

There, one observes emission from both a PDR and from shocked gas and both processes will have to be considered a priori. An overview of the region near the binary Class\,0 source VLA\,1623 \citep{looney2000} in the \roa\ cloud core, toward which spectral line data of both \molh\ and oxygen exist, is shown in Fig.\,\ref{VLA}. This study focusses on the analysis of these data with the aim to obtain the oxygen abundance in the \roca.

This paper is organized as follows: In Sect.\,2, we report on the ISOCAM observations, account for their reduction and briefly present the main results. In this section, we also give an account of the LWS observations and describe the results in considerable detail. In Sect.\,3, we derive the temperature and the column density of the emitting \molh\ gas and present our results for the accompanying oxygen gas, with emphasis on the elemental abundance. Finally, in Sect.\,4, we briefly summarize our main conclusions from this work. 

\begin{table}[h]
\begin{flushleft}
 \caption{\label{PAH_width} The peak position and the width of the Lorentzian profiles
used to fit the PAH features.}
\resizebox{\hsize}{!}{
\begin{tabular}{ccc}
\hline 
\noalign{\smallskip}
PAH feature (\um) & Peak position (cm$^{-1}$) &  FWHM  (cm$^{-1}$)    \\
\noalign{\smallskip}
\hline
  \hline
  \noalign{\smallskip}
 \noalign{\smallskip}
6.2         & 1596.9                   & 70  \\
7.7         & 1295.8                   & 113 \\
8.6         & 1176.5                   & 40  \\
11.3        & 877.5                    & 25  \\
12.7        & 781.2 - 800              & 40  \\
  \noalign{\smallskip}
  \hline
  \end{tabular}
  }
\end{flushleft} 
\end{table}

\section{Observations, data reductions and results}

\subsection{ISO-CAM CVF}

The pipeline processed ISOCAM-CVF data of \roa\ were extracted from the ISO archive (TDT numbers 4520111 and 45601809), covering a region of $3^{\prime} \times 3^{\prime}$. These observations trace the CO outflow from VLA\,1623 \citep[e.g.,][]{andre1990,dent1995} in a map of $32 \times 32$ pixels, each representing 6$^{\prime\prime} \times 6^{\prime\prime}$ (the angular resolution of ISO-CAM is diffraction limited). In each pixel of the image frames, the light has been dispersed by the Circular Variable Filter (CVF) to produce an infrared spectrum, so that the CAM data thus are represented by data cubes with two spatial and one spectral dimension. In addition, two wavelength channels result in frames, in which each pixel contains a short-wavelength (SW, $5-9.5 {\mu}$m) and a long-wavelength (LW, $9.5-16.5 {\mu}$m) spectrum and the resolution is $\lambda/\Delta \lambda = 34-52$ \citep{blommaert2003}. Each pixel had both of its spectra examined individually and reduced manually as described below.

Using known point sources in the region for positional reference, it can be seen that the SW and LW frames are shifted by 3 pixels in the $y$-direction. With a `dead row', in each of the SW and LW frames, there are $\sim 850$ spectra which cover the full $5-16.5\,{\mu}$m range. The spectra of the two wavelength regions were joined together by adjusting either the SW or LW part vertically. The relative rotation was within a small fraction of a pixel and we did not attempt to correct for that\footnote{More specifically, the equatorial J2000.0 coordinates for the SW-frame (TDT 4520111) are RA = \radot{16}{26}{21}{386}, Dec = \decdot{$-24$}{23}{59}{28} at spacecraft roll angle \adegdot{98}{31928}. For the LW-frame (TDT 45601809) the corresponding data are RA = \radot{16}{26}{21}{259} and Dec = \decdot{$-24$}{24}{11}{16} at roll = \adegdot{98}{51982}. The coordinates refer to the respective map centres.}.

\begin{figure*}[ht]
  \resizebox{\hsize}{!}{
  \rotatebox{270}{\includegraphics{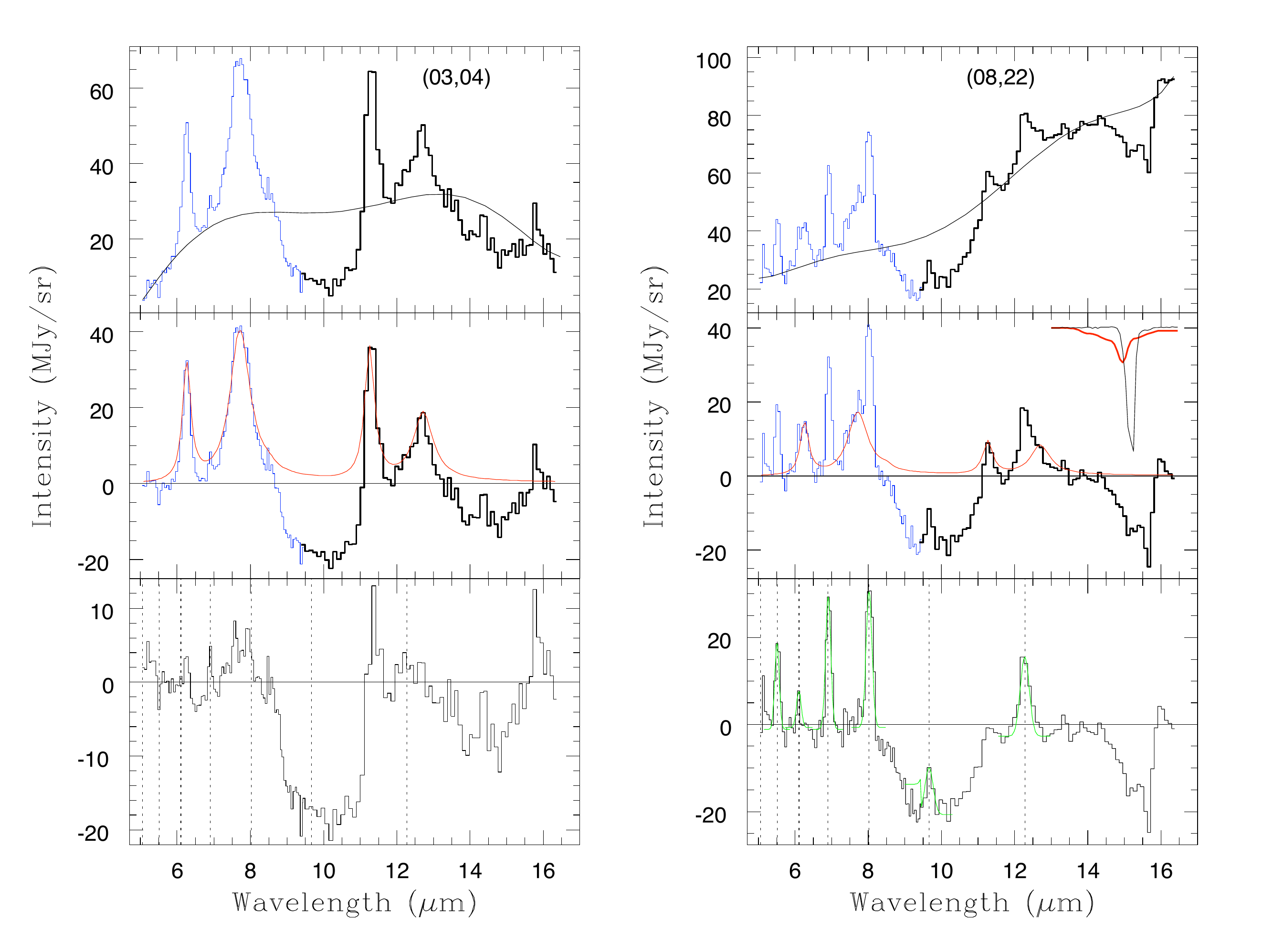}}
  }
  \caption{{\bf Left:} The observed CVF-spectrum in CAM-pixel (3, 4), where PAH features dominate the spectrum. The short-wave and long-wave bands are shown as histograms with thin and thick lines, respectively. Fits to the continuum and the PAH features with Lorentzian profiles are shown in the upper two frames. In the lower frame, these components have been subtracted and the wavelengths of \molh\ lines are shown by vertical dashed lines for reference. 
{\bf Right:} The spectrum of pixel (8, 22) corresponds to HH\,313\,A and contains less dominant PAH-emission. Instead, a number of \molh\ lines are discernable. The upper two frames show again the fits to the continuum and PAHs, respectively, whereas the lower frame reveals the rotational line spectrum of molecular hydrogen. In addition to the emission features, there are also absorption bands from solids, viz. due to silicates at 10\,\um\ and due to CO$_2$ at 15\,\um. Theoretical models of both gaseous (red) and solid (black) CO$_2$ absorption, at the appropriate spectral resolution of the CAM-CVF but arbitrarily shifted in the $y$-direction, are shown in the upper right corner of the middle frame.
  }
  \label{spectrum}
\end{figure*}

\subsubsection{Solid state and PAH features}

Each of these spectra shows presence of the silicate dust absorption at 10\,${\mu}$m, along with very strong emission at  6.2, 7.7, 11.3 and 
12.7 ${\mu}$m thought to be due to  polycyclic aromatic hydrocarbons \citep[PAHs,][]{leger1984,allamandola1985}. In order to extract the emission due to PAHs, we fitted a continuum below the PAH features and over the silicate absorption (Fig.\,\ref{spectrum}). After subtracting the continuum, we fitted the PAH emission, assuming that each line profile can be described by a Lorentzian profile \citep{boulanger2005}. The peak position and the width of each PAH feature are listed in Table\,\ref{PAH_width}.

The star S\,1 is illuminating the north-eastern part of \roa, while HD\,147889 is heating the dust in the south-western region of the ISOCAM map (Fig.\,\ref{VLA}). The ratio of the PAH bands at 7.7\,\um\ and 12.7\,${\mu}$m can be used as an indicator of the degree of ionization of the PAH molecules and by comparing the maps of these two bands, it can be seen that the FUV photons from the early B-type star HD\,147889 can penetrate the cloud and ionize the PAH molecules more effectively than those from S\,1, a star of slightly later spectral type. This is entirely in line with the conclusions reached by \citet{liseau1999} on the basis of their ISO-LWS observations.

The data for the PAHs and the solid state features in \roa\ have been presented and discussed by \citet{alexander2003}. For the ice-ratio CO$_2$/\water, these authors give a range of $0 - 0.4$, but left unspecified which region in \ro\ these numbers do refer to. In a limited number of pixels along the outflow, we detect a sharp absorption feature due to CO$_{2}$ ice. At the resolution of the ISOCAM-CVF, it is not possible to make a detailed study of the matrix of the ice, i.e., whether it is pure CO$_{2}$ ice or mixed with water. However, we do not detect water-ice absorption at 6.2\,${\mu}$m as any absorption would be filled in by the PAH 6.2\,\um\ emission feature. Also the broad libration band of water-ice would suffer from contamination by PAH emission at 11.3\,\um\ and 12.7\,${\mu}$m.

The broad absorption wing near 15\,\um\ could be indicative of CO$_2$ gas and model fitting would require likely unrealistically large column densities (Fig.\,\ref{spectrum}). This would be in large contrast to what has been observed elsewhere at much higher spectral resolution \citep{EvD1998}. Therefore, mainly due to the difficulty to correctly identify the continuum level we are unable to make any quantitative assessment of the column density of CO$_2$ gas. The situation seems better for the ice feature though, for which we estimate $N$(CO$_2)_{\rm ice}=(1.4 \pm 0.4) \times 10^{17}$\,\cmtwo, and where we have used the absorption cross section provided by \citet{gerakines1995}. This particular estimate refers to the pixel (8, 22), see Fig.\,\ref{spectrum}.
 
\begin{table*}[ht]
\begin{flushleft}
 \caption{\label{LWS_obs} ISO-LWS [O\,I] line fluxes and line ratios for the VLA\,1623 outflow and the neighbouring \ro\ PDR}
\resizebox{\hsize}{!}{
\begin{tabular}{lllllllll}
\hline 
\noalign{\smallskip}
TDT        & RA(J2000)      & Dec(J2000)                & \powten{12}  $F_{63\,\mu {\rm m} }$    &  S/N$_{63}$  & \powten{12}  $F_{145\,\mu {\rm m} }$  & S/N$_{145}$  & Project Name/ & Comment   \\             
number  & (h   m   s)         & (\adeg\     \amin\     \asec) &  (\ecs)                                                 &                         &  (\ecs)                                                        &                          &  Position           &                       \\     
\noalign{\smallskip}
\hline
  \hline
  \noalign{\smallskip}
{\bf Flow}       &                             &                        &                                  &                       &                                  &           &                                      &                       \\
29200533$^a$ & 16 26 26.27      &  $-24$ 24 30 & $18.52 \pm 1.29$ &     14             &  $4.47  \pm 1.50$  &   2.6  &    VLA 1623                & Origin (flow 0\asec\ offset)   \\			  
29200534$^b$ & 16 26 17.22      & $-24$ 23 06  & $22.54 \pm 0.80$ &     28             &  $4.22  \pm 0.86$  &   4.9  &    VLA 1623 flow(1,1) &  150\asec\ NW  (pa=304\adeg)\\
29200534          & 16 26 23.25      & $-24$ 24 02  & $18.52 \pm 1.27$ &     15             &  $4.47  \pm 1.35$  &   3.3  &    VLA 1623 flow(2,1) &  \phantom{1}50\asec\ NW (pa=304\adeg)\\
29200534          & 16 26 29.29      & $-24$ 24 58  & $21.92 \pm 1.00$ &     22             &  $8.94  \pm 1.62$  &   5.5  &    VLA 1623 flow(3,1) & \phantom{1}50\asec\ SE   (pa=124\adeg) \\
\noalign{\smallskip}
\hline
  \noalign{\smallskip}  
{\bf PDR}   &                             &                        &                                  &         &                                   &        &                                     &                        \\
45400801 & 16 25 55.56      & $-24$ 25 39 & $10.08  \pm 0.77$ &   13  &  $4.02  \pm 0.80$   & 5.0 & ROPH\_EW ew2      &  \about south 1\amin\ west 7\amin \\
45400801 & 16 26 08.74      & $-24$ 25 40 & $12.66  \pm 0.62$ &    20 &  $4.92  \pm 1.09$   & 4.5 &  ROPH\_EW ew3     &  \about south 1\amin\ west 4\amin \\          
45400801 & 16 26 21.92      & $-24$ 25 40 & $10.56  \pm 0.81$ &   13  &  $3.45  \pm  0.60$  & 5.8 & ROPH\_EW ew4      &  \about south 1\amin\ west 1\amin \\          
45400801 & 16 26 35.10      & $-24$ 25 41 & $12.66  \pm 0.36$ &    35 &  $4.41  \pm 0.86$   &  5.1 &  ROPH\_EW ew5    &  \about south 1\amin\ east 2\amin \\
{\bf PDR$_{\rm ave}$}$^c$       &      &                         & $11.49  \pm 0.64$ &   18 &  $4.20 \pm  0.86$   & 4.9 &                                     &    \\   
   \noalign{\smallskip}      
  \hline
 \noalign{\smallskip}   
{\bf Offset}$^d$ &16 26 35.32 & $-24$ 25 54  & $\phantom{1}9.88 \pm 1.13$ & \phantom{1}9 & $3.96 \pm 0.93$   &   4.2  &    VLA 1623 flow(4,1) & 150\asec\ SE (pa=124\adeg) \\
\noalign{\smallskip} 
\hline
  \noalign{\smallskip}            
 \noalign{\smallskip} 
{\bf Flow$-$PDR$_{\rm ave}$}&  &                &   11.05                    &          &   0.02                        &          &   552 = Line Ratio     &   150\asec\ NW (pa=304\adeg)    \\
                    &                            &                         &  \phantom{1}7.03 &          &   0.27                        &          &   \phantom{1}26        &   \phantom{1}50\asec\ NW  (pa=304\adeg)  \\  
                    &                            &                         &  \phantom{1}7.96 &          &   0.27                        &          &   \phantom{1}29        &  Origin (flow 0\asec\ offset)    \\      
{\bf Flow$-$Offset} &              &                         & 12.66                      &          &  0.26                        &           &    \phantom{1}50 =  Line Ratio       &  150\asec\ NW (pa=304\adeg)\\
                             &                   &                         &  \phantom{1}8.64 &          &  0.48                        &           &   \phantom{1}18        &  \phantom{1}50\asec\ NW (pa=304\adeg)\\   
                             &                   &                         &  \phantom{1}8.64 &          &  0.51                        &           &   \phantom{1}17        & Origin (flow 0\asec\ offset)   \\	
 \noalign{\smallskip}
  \hline
\noalign{\smallskip}
  
{\bf $<$Flow-PDR$_{\rm ave}$$>$}  &   &  &  \phantom{1}8.68  &            &  0.27                        &           &  $>25$                         &        \\              
{\bf $<$Flow$-$Offset$>$}  &    &                  &   \phantom{1}9.98 &            & 0.41                         &          &   \phantom{1}24          &                \\
 \noalign{\smallskip}
  \hline
  \end{tabular}
    }
\end{flushleft}
  Notes to the Table: \\
 $^a$  Center position related to the strip map (TDT 29200534), \\
 $^b$ which is  along position angle 124\adeg\ ($\pm 180$\adeg) with 150\asec\ spacings of 4 individual pointings. \\
  $^c$ The surrounding PDR provides one type of background estimate. \\
  $^d$ TDT 29200534: This position of the strip scan is outside the flow, yet close enough to serve as another reference position for background estimation. 
\end{table*}

\subsubsection{\molh-lines}

The residuals after fitting the PAHs show that in several pixels, there is a series of emission lines due to pure rotational transitions of molecular hydrogen, H$_{2}$. The combination of SW and LW spectra made it possible for us to detect the transitions from S\,(2) up to S\,(7). These lines are not resolved at the resolution of the ISOCAM-CVF and were fitted using Gaussian profiles in order to estimate the line intensities (bottom right panel in Fig.\,\ref{spectrum}). There are no detections of possible other line emitters, such as, for instance, [Ne II]\,12.8\,\um.

\begin{figure}[ht]
  \resizebox{\hsize}{!}{
  \rotatebox{00}{\includegraphics{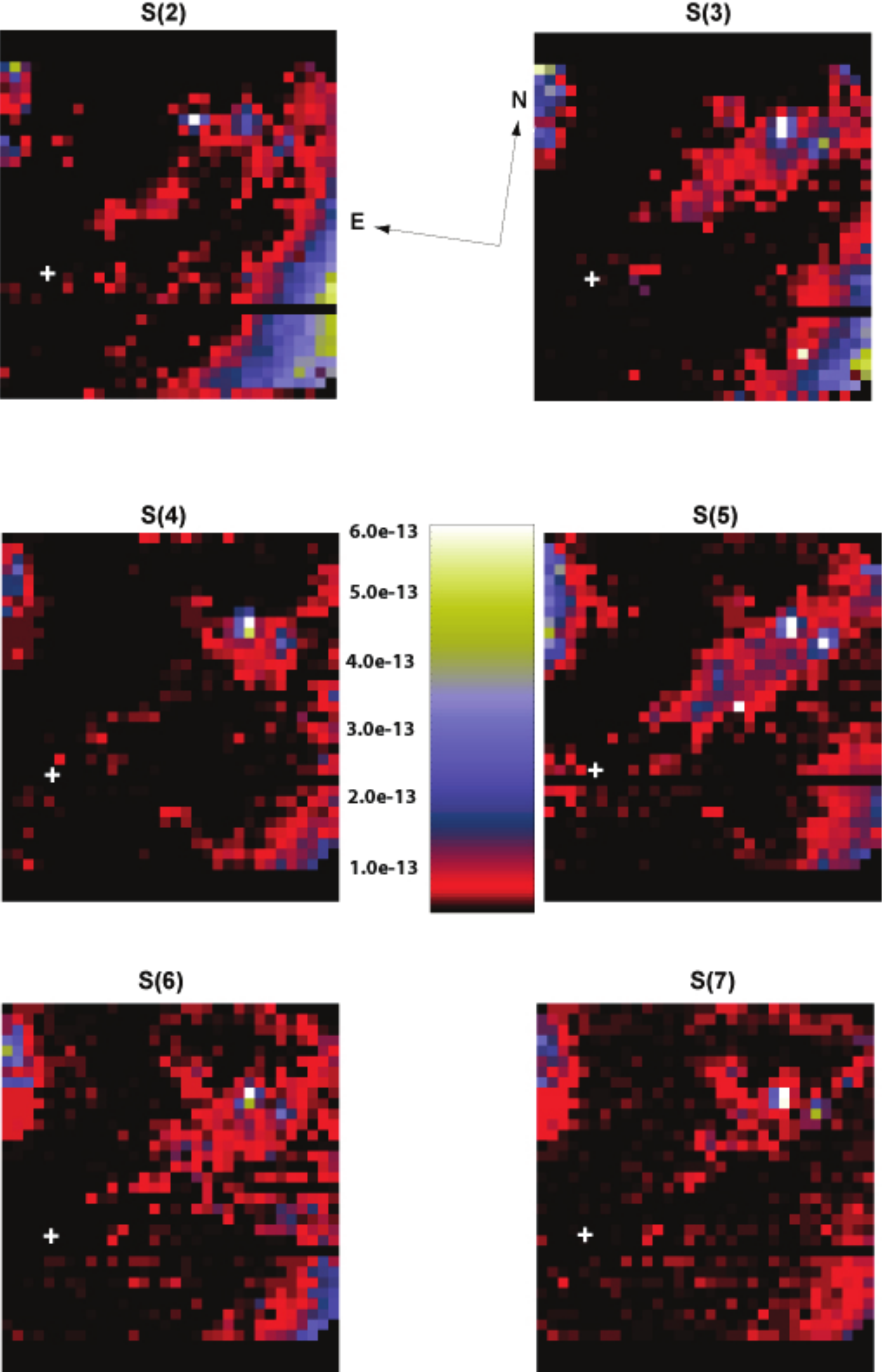}}
  }
  \caption{For convenience, the ISOCAM-CVF frames have been rotated (cf. Fig.\,\ref{VLA}) and arranged so that the lines originating from the para-states are shown in the panels to the left and those from the ortho-states to the right. The dark horizontal bar is due to a row of reduced pixel sensitivity, whereas the dark edges are due to loss of data by the alignment procedure. 
The image dimensions are $192^{\prime \prime} \times 210^{\prime \prime}$ in the horizontal and vertical direction, respectively. In the lower right corner, the PDR is clearly seen, especially prominently in the S\,(2) line. The scale bar is in units of \ecs\ and the brightest spot is HH\,313\,A, the proper motion of which associates it with the CO outflow. To its right is HH\,313\,B, which is likely unrelated to VLA\,1623 but rather emanates from GSS\,30 \citep{caratti2006}. The position of the outflow source VLA\,1623 is indicated by a white cross.
  }
  \label{h2_lines}
\end{figure}

The spatial distribution of the individual \molh\ lines is displayed in Fig.\,\ref{h2_lines}. \molh\ line emission associated with the outflow from VLA\,1623 is clearly seen. Along this flow, the emission is particularly prominent in the S\,(5) and S\,(6) lines. This indicates that the \molh\ outflow gas is in a state of elevated excitation and that temperatures are likely to be relatively high. In contrast, the PDR emission is especially prominent in the low-excitation S\,(2) emission, suggesting this gas to be at lower temperatures, consistent with the results from theoretical models of the \roa\ PDR \citep{liseau1999,spaans2001,habart2003,kulesa2005,hollenbach2009}. It should be noted that the sparse distribution of the S\,(4) line is due mainly to the strong contamination by the 7.7\,\um\ PAH feature. The bright pixels in the upper right correspond to the known Herbig-Haro objects HH\,313\,A and B \citep[e.g.,][]{gomez2003}.

\subsection{ISO-LWS and [O\,I] lines}

The oxygen spectral line data analysed in this paper had been obtained with the Long Wavelength Spectrometer (LWS) on board ISO. In Table\,\ref{LWS_obs}, the observations which were retrieved from the archive are identified by their TDT numbers. The presented data are pipeline reduced and line fluxes were extracted from the detectors SW\,3 for the \oishort\ line and LW\,3 for the \oilong\ line, respectively. To estimate the flux, we fitted the line after continuum subtraction with a Gaussian,  the width of which was kept fixed to the resolution of the short- and long-wavelength LWS bands, i.e., 0.29 and 0.6\,\um, respectively. The conservative error estimate reflects the peak-to-peak of the noise and the absolute accuracy of the LWS data is better than 30\%. At these wavelengths, extended source corrections result in effective LWS-beams of $1.41 \times 10^{-7}$\,sr and  $9.35 \times 10^{-8}$\,sr, respectively \citep{gry2003}. 

The TDT 29200534 data set contains a strip scan along the position angle 124\adeg\ and with 100\asec\ spacings. The center coordinates of the scan are given by TDT 29200533, i.e. toward VLA\,1623. The pointings are along the CO outflow from that source. The data of the northwest flow provide the basis for our discussion below, whereas the observations toward the offset 150\asec\ southeast provided a (quasi)simultaneous measurement of the background. This position, named ``Offset" in Table\,\ref{LWS_obs} lies outside the CO outflow, which is changing its direction, yet it is sufficiently close to  serve as a reference position for the flow data in the northwest. 

Although the use of a single reference position is common practice when observing dark clouds, we used also another set of observations, i.e. those of the surrounding PDR, to gauge the [O\,I] emission from and near the VLA\,1623 outflow. The comparison with this statistically estimated background should provide us with the means to assess the accuracy of the results (cf. Table\,\ref{OIlines}). These data showed that the PDR provides a background of essentially constant intensity and, consistent with the other data set, that the 63\,\um\ emission along the flow is enhanced by a factor of almost two, while the \oilong\ line resembles the background.

For both data sets, the signal-to-noise ratio (S/N) of the 63\,\um\ data is generally much higher than 10, whereas that of the \oilong\ lines is roughly 3 to 5. Subtracting the Offset and averaged PDR fluxes from the flow data revealed comparable \oishort\ excesses in both cases, but led to small residuals at 145\,\um, resulting in large relative line ratios, in excess of about 20 (see the last two rows in Table\,\ref{LWS_obs}). 

The \oilong\ datum for the 50\asec\ southeast position shows an exceptionally large flux value. This cannot be explained in terms of contamination by the CO\,($J$=18-17) line at 144.8\,\um, as no other CO emission is detected elsewhere in the spectrum (similarly, any potential blending of the \oishort\ line with \water\,$(8_{08}-7_{17})$ can generally be dismissed on similar grounds). The continuum level is also enhanced, by a factor of 3 to 4, whereas the \oishort\ emission is of the same order as the other flow values (Table\,\ref{LWS_obs}). However, this position is outside the CAM frame containing the \molh\ map, and it is this map which is at the focus of the following discussion. 

\section{Discussion}

\subsection{Extinction of the \molh\ lines}

In dense clouds, corrections for extinction by dust might become necessary even in the infrared (Appendix\,A). The CVF admits the three para-lines S(2), S(4) and S(6) at 12.3, 8.0 and 6.1\,\um, respectively, and the three ortho-lines S(3), S(5) and S(7) at the respective wavelengths of 9.7, 6.9 and 5.5\,\um. The integer in parentheses refers to the rotational quantum number of the lower state, $J_{\rm low}=J_{\rm up}-2$. The location of the lines in the CAM-CVF spectra is shown in Fig.\,\ref{spectrum}. 

\begin{figure}[ht]
  \resizebox{\hsize}{!}{
  \rotatebox{00}{\includegraphics{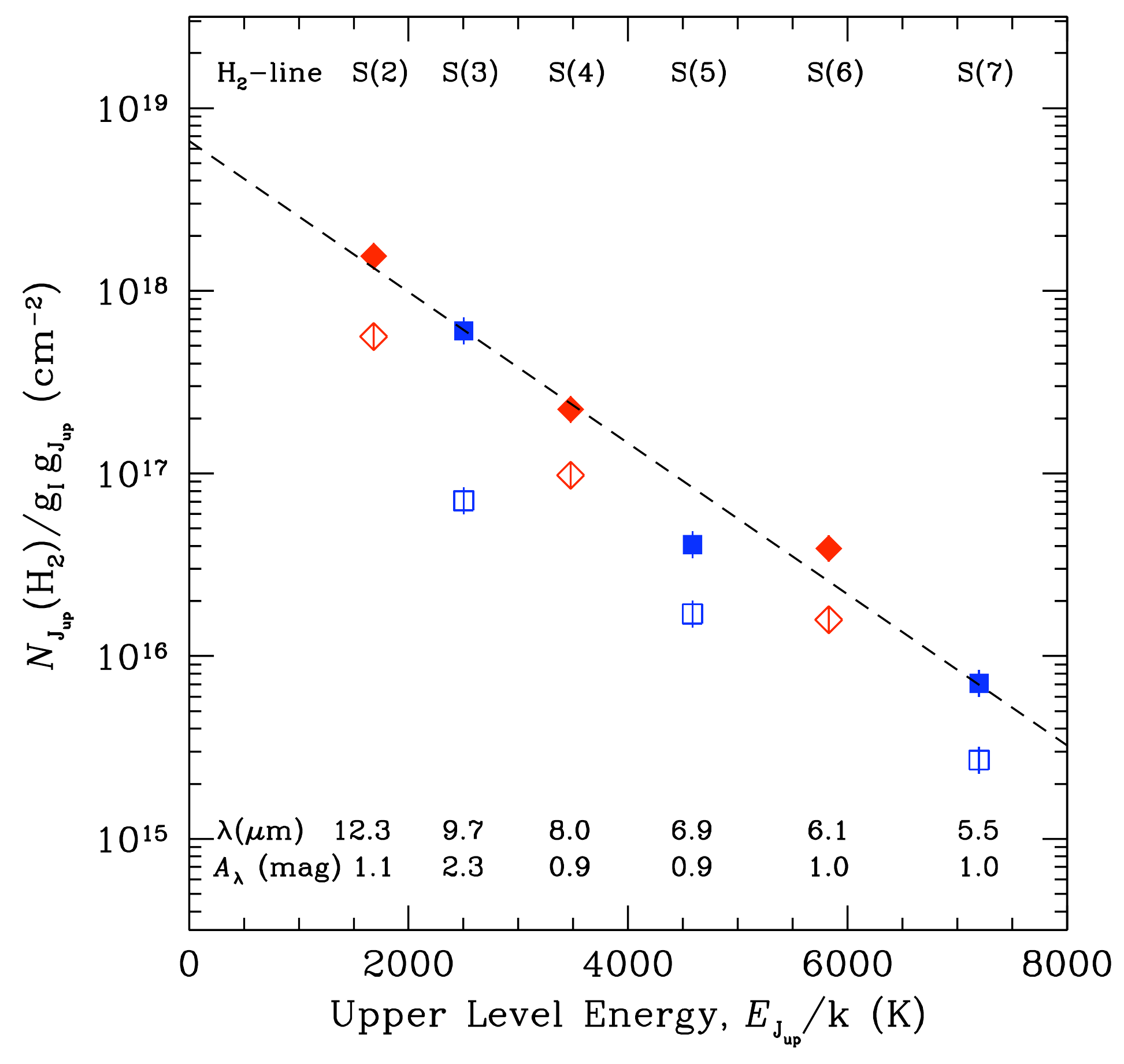}}
  }
  \caption{The rotational \molh-diagram for CAM pixel (8, 23), HH\,313\,A. At their respective upper level energy (in K), the transitions are identified. Points with error bars (para: red diamonds, ortho: blue squares) refer to observed (open symbols) and extinction corrected values (filled), respectively. The value of the visual extinction, derived from $\chi^2$-minimization, is \av\,=\,28\,mag and the corresponding $A_{\lambda}$ in magnitudes are given below the graph. The inverse slope of the straight line is a measure of the gas temperature, $T=1050 \pm 150$\,K. At this particular location in the outflow from VLA\,1623, the derived ortho-to-para ratio is ${\rm {o/p}}=2.1 \pm 0.2$.
  }
  \label{rot_diag}
\end{figure}

\begin{figure}[ht]
  \resizebox{\hsize}{!}{
  \rotatebox{00}{\includegraphics{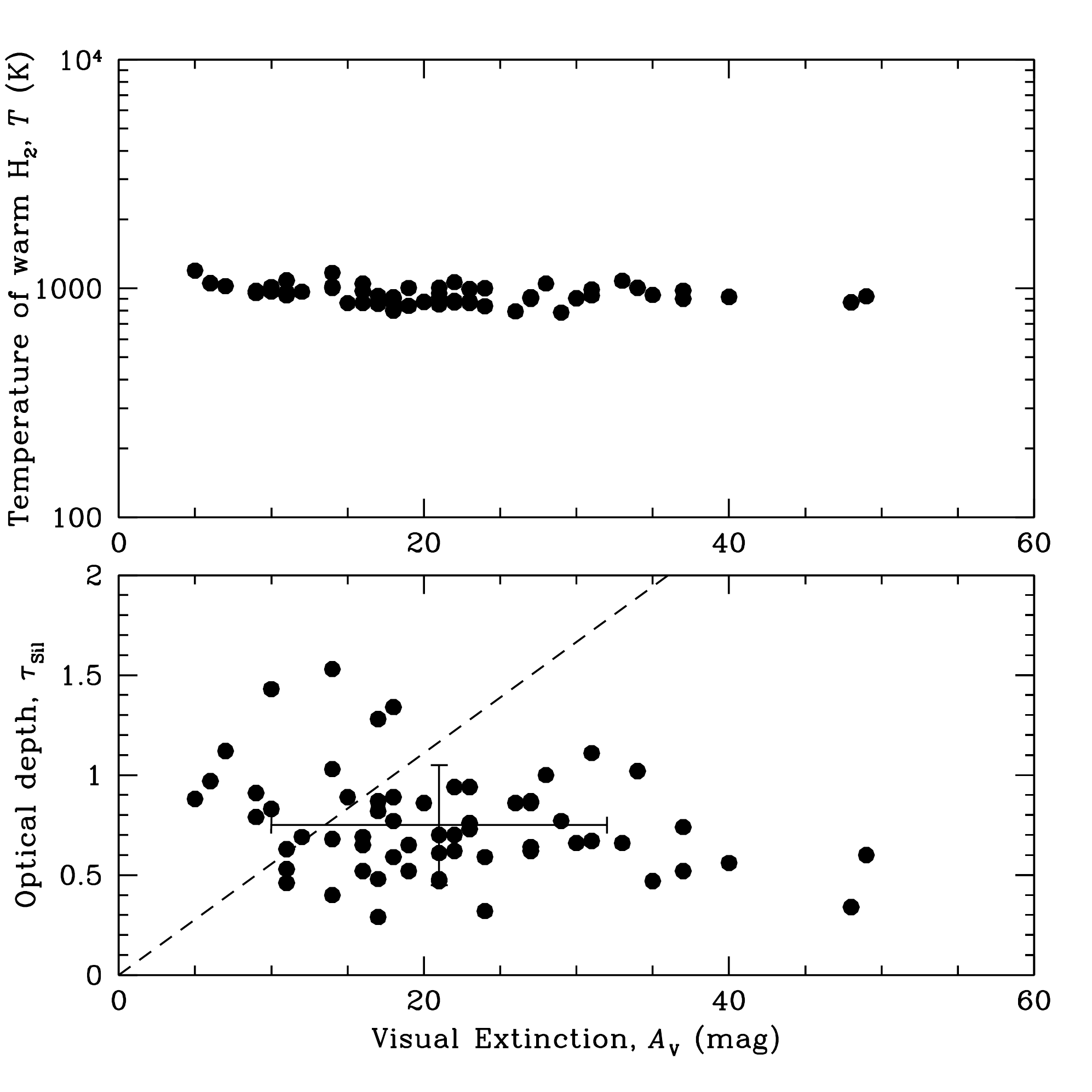}}
  }
  \caption{{\bf Top:} Visual extinction, \av, and temperature of the \molh-gas, showing no dependence as expected. 
          {\bf Bottom:} Visual extinction, \av, and optical depth in the silicate absorption feature, $\tau_{\rm Sil}$. The error symbol shows the average of the points and their standard deviation. The dashed line displays the relation for diffuse clouds of \citet{whittet2008}.
  }
  \label{T_AV_Sil}
\end{figure}

The S(3) line is coinciding in wavelength with the broad 10\,\um-silicate feature and is therefore most sensitive to the dust extinction. In the observed CAM-CVF frame, both high and low values of \av\ are derived (see Figs.\,\ref{rot_diag} and \ref{T_AV_Sil}), with an average of $21 \pm 10$ magnitudes over 63 pixels, in which we detected six \molh\ lines. For a larger number of pixels, i.e., 162 pixels with 5 detected lines, this value is not significantly lower, i.e. $16 \pm 11$\,mag. Partially responsible for the large spread might be the fact that the actual extinction curve is different from the one used here.

Extinction values for extended regions of the \roc\ and as high as those derived here have been reported also by others \citep[e.g., ][]{davis1999,whittet2008}. Especially remarkable, and in line with our own findings, is the result of \citet{davis1999} for HH\,313A, who found at \asecdot{0}{6} resolution variations of the K-band extinction by 3 magnitudes ($\Delta$\av=30\,mag) over only 6\asec, i.e., the size of one CAM pixel. Obviously, the surface layers of the \roa\ cloud are extremely inhomogeneous. In fact, this is evident already in the original of Fig.\,\ref{VLA}, which reveals the large increase of the extinction even longward of 3\,\um\ and also clearly shows the dust distribution over the flow region to be patchy and filamentary.

The observed optical depth of the silicate absorption feature, $\tau_{\rm Sil}$, could potentially be used as an independent way of estimating the extinction \citep[e.g.,][and references therein]{whittet2008}. In the lower panel of Fig.\,\ref{T_AV_Sil}, a plot of $\tau_{\rm Sil}$ versus \av\ is shown. Evidently, no close correlation, such as that shown by the straight line and valid for diffuse clouds, does exist for the dense core \roa. From the work by \citet{chiar2007} it is evident that in dark clouds, this relationship does generally not hold for optical depths larger than about 0.6, corresponding to \av\,\gapprox\,10\,mag. In our work, $\tau_{\rm Sil}$ generally exceeds 0.6 (Fig.\,\ref{T_AV_Sil}) and we decided against the use of the diffuse cloud relation.

\subsection{Physical parameters of the \molh\ gas}

The \molh\ parameters presented here are based on a `rotation diagram' analysis, in which it is assumed that the emitting gas is in local thermal equilibrium (LTE) at a single gas kinetic temperature and that the emission is optically thin (line center opacity much less than unity). In Appendix\,B, the method and the validity of the assumptions are examined. 

The \molh\ gas associated with the VLA\,1623 outflow is remarkably isothermal, $T=10^3$\, K (formally $962 \pm 129$\,K, cf. top panel of Fig.\,\ref{T_AV_Sil}) and the average column density of the warm gas is $N({\rm H}_2) = 3.5 \times 10^{19}$\,\cmtwo.  The ortho-to-para ratio along the flow is determined as $2.2 \pm 0.4$, respectively.

The mass of warm \molh-gas along the outflow from VLA\,1623 amounts to $\ge 8 \times 10^{-4}$\,\msun, which radiates more than $7 \times 10^{-2}$\,\lsun. This mass is nearly comparable (\about\,50\%) 
to that of the cold gas traced in CO in the northwestern outflow as derived by \citet{andre1990}, assuming a 25\% smaller distance \citep[120\,pc:][]{lombardi2008,snow2008}. These estimates exclude the confusing blueshifted flow from SM\,1N  \citep{narayanan2006}.

\begin{figure}[ht]
  \resizebox{\hsize}{!}{
  \rotatebox{00}{\includegraphics{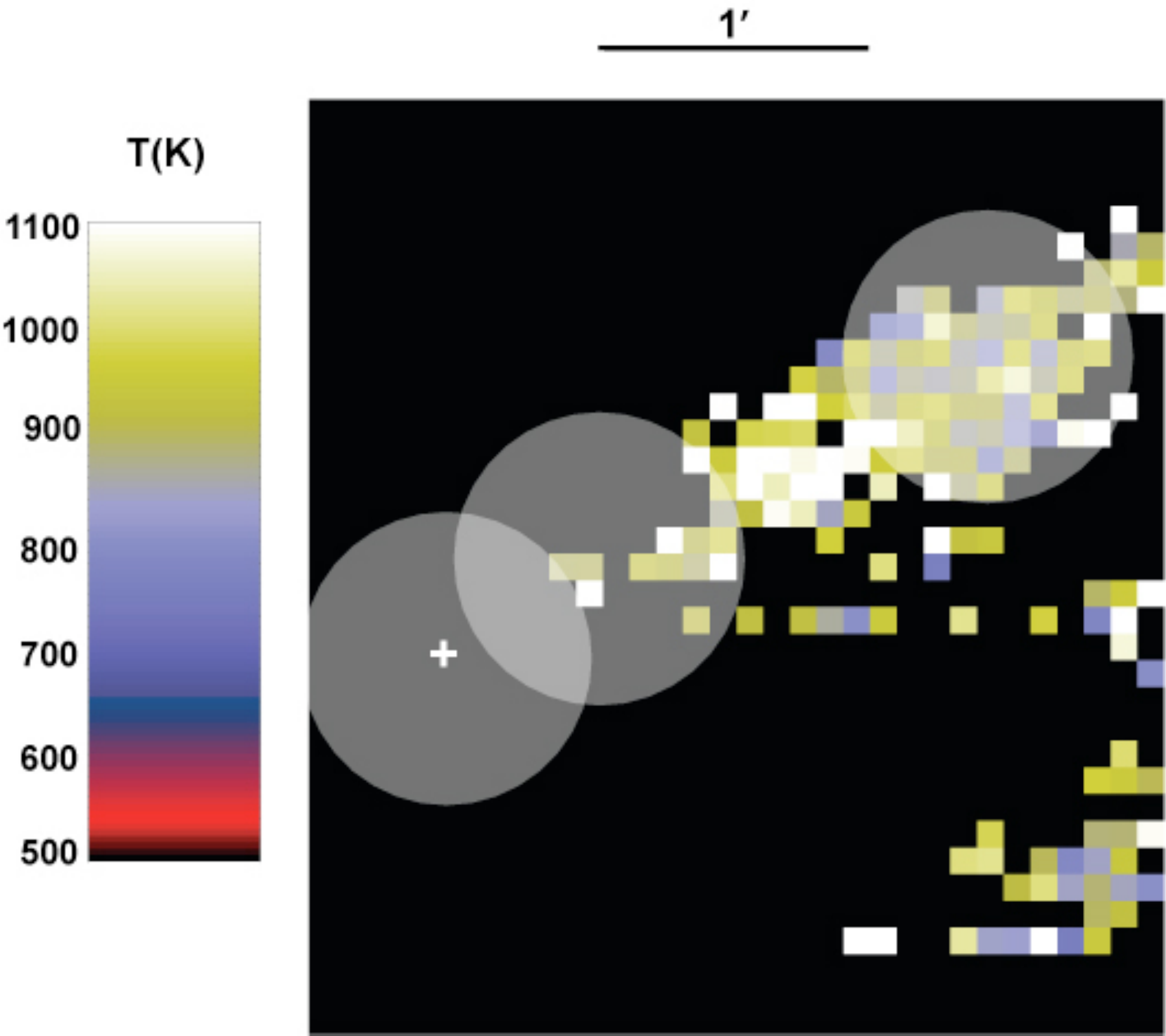}}
  }
  \caption{The spatial distribution of the kinetic gas temperature derived from the \molh\ S\,(2) to S\,(6) lines. The position of the outflow source VLA\,1623 is indicated by a white cross and the LWS pointings are shown by circular beams of diameter 70\asec\ and their positions refer to 0\asec, 50\asec\ NW and 150\asec\ NW (see Table\,\ref{LWS_obs}).
  }
  \label{h2_temp}
\end{figure}

\begin{table*}[ht]
\begin{flushleft}
 \caption{\label{OIlines} Predicted [O\,I] $^3$P fluxes for the ISO-LWS at position 150\asec\,NW, assuming [O/H]\,$= 3.4 \times 10^{-4}$.} 
\resizebox{\hsize}{!}{
\begin{tabular}{rllrclclccc}
\hline 
\noalign{\smallskip}
 $T$  &  $N({\rm H}_2)$             & $N({\rm O})$                 & $n$(\molh)                         & $\tau_{63\,\mu{\rm m}}$ & $F_{63\,\mu{\rm m}}$     & $\tau_{145\,\mu{\rm m}}$ & $F_{145\,\mu{\rm m}}$ & $F_{63}/F_{145}$ & $F_{\rm obs}/F_{\rm pred}$ & $F_{\rm obs}/F_{\rm pred}$  \\
 (K)   &   (\cmtwo)                         &  (\cmtwo)                       &  (\cmthree)                         &                                             & (\ecs)                                  &                                              &  (\ecs)                    & pred$^a$ & 63\,\um$^b$   &145\,\um$^b$ \\
\noalign{\smallskip}
\hline
  \hline
  \noalign{\smallskip}
 \noalign{\smallskip}
 20    & $ 3.0 \times 10^{22}$   &  $ 2.0 \times 10^{19}$  &  $ 3 \times 10^{5}$ & 110   &  $ 4.2 \times 10^{-14}$  &   $9.7 \times 10^{-4}$     &  $ 4.5 \times 10^{-16}$  &   92                           &     262           &44\\               
 \noalign{\smallskip}                     
1000 & $ 3.5 \times 10^{19}$  &  $ 2.4 \times 10^{16}$  &  $ 5 \times 10^{3}$ & 0.12  &  $ 8.9 \times 10^{-12}$  &       $-0.06$                         & $ 8.0 \times 10^{-13}$   &   11                         &     1.1           & 0.025 \\                              
          &                                         &                                          &  $ 3 \times 10^{5}$ & 0.05   &  $ 1.5 \times 10^{-10}$ &       $-0.01$                         & $ 4.5 \times 10^{-12}$   &    33                         &     0.07        & 0.004\\                               
 \noalign{\smallskip}
  \hline
  \end{tabular}
  }
\end{flushleft}
Notes to the Table: $^a$ Observed ratio $F_{63}/F_{145} \sim 50$ (see Table\,\ref{LWS_obs}). $^b$ Observed fluxes $F_{63}=1 \times 10^{-11}$ and $F_{145}=2 \times 10^{-14}$\,\ecs.
\end{table*}

\begin{figure}[ht]
  \resizebox{\hsize}{!}{
  \rotatebox{00}{\includegraphics{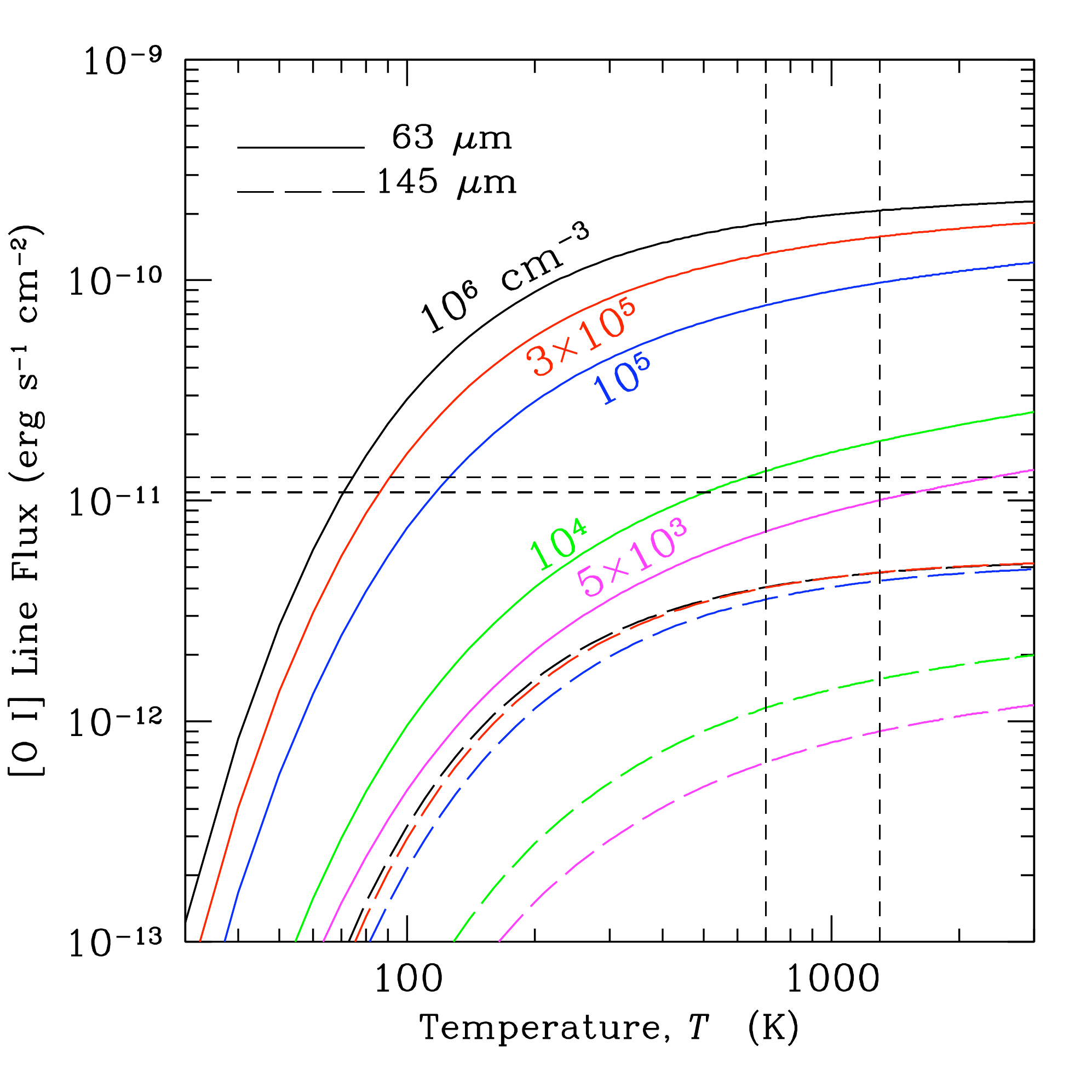}}
  }
  \caption{The predicted flux into the LWS beam of \oishort\ (solid lines) and \oilong\ (long dashes) is shown as function of the temperature and for the indicated values of the \molh\ density, assuming an oxygen 
  abundance of  $3.4 \times 10^{-4}$ relative to H and the column density determined from the rotational \molh\ observations, i.e. $N({\rm H}_2)=3.5 \times 10^{19}$\,\cmtwo. Values observed at the VLA\,1623 outflow
   position 150\asec\ NW are found inside the dashed lines and bound by the $\pm 2 \sigma$ values of the \oishort\ line flux and of the temperature. The upper level energies for the \oishort\ and the  \oilong\ lines are
    228\,K and 326\,K, respectively. 
  }
  \label{OI_flux}
\end{figure}
\begin{figure}[ht]
  \resizebox{\hsize}{!}{
  \rotatebox{00}{\includegraphics{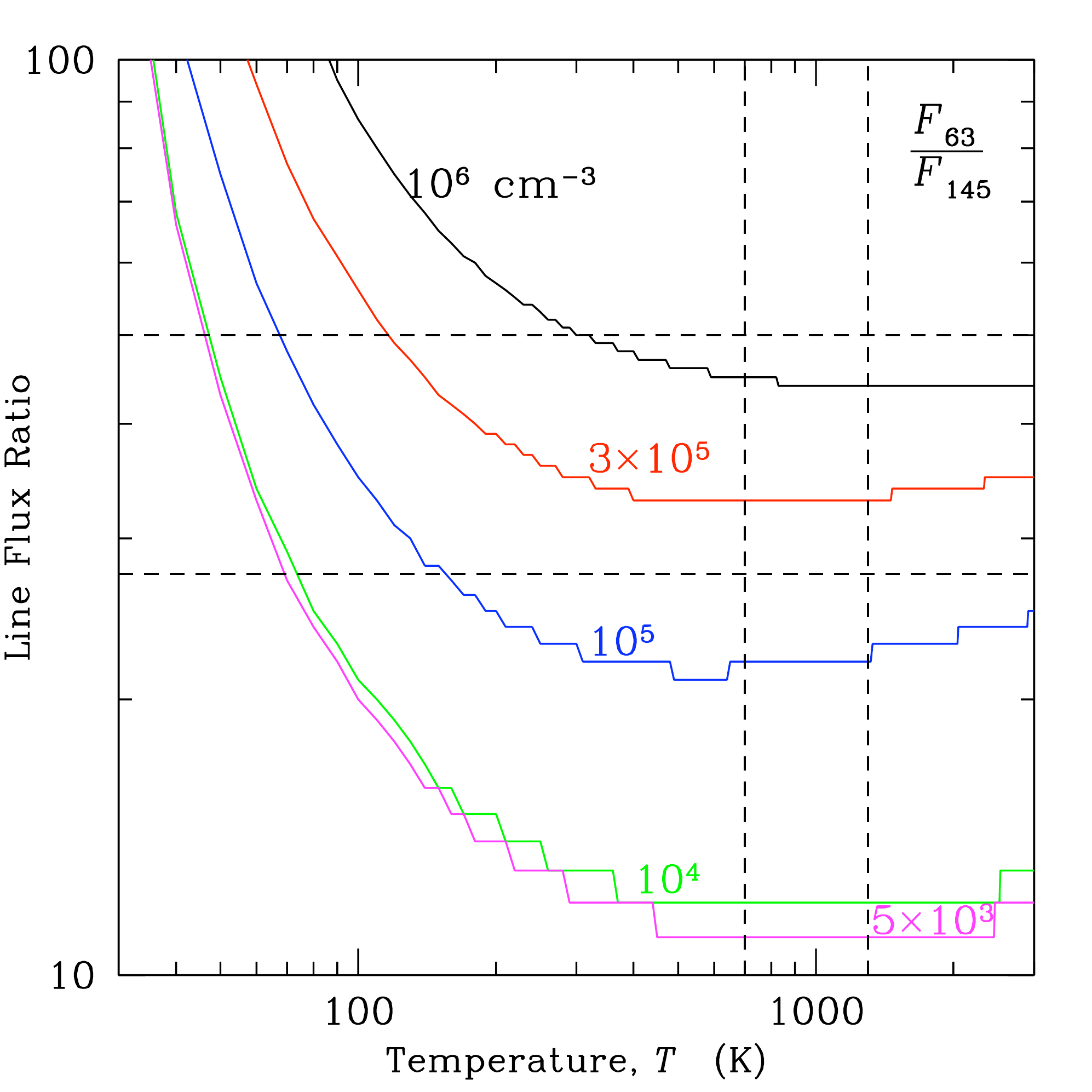}}
  }
  \caption{The line flux ratios of the  \oishort\ and \oilong\ lines are shown as a function of the temperature for five values of the \molh\ density. The slightly different beam sizes of the LWS have been taken into account 
  and the computations assumed the common line width of 1\,\kms. Values observed along the VLA\,1623 outflow are found inside the dashed lines and bound by the $\pm 2 \sigma$ values ($\pm 300$\,K) of the
   temperature.
    }
  \label{OI_ratio}
\end{figure}

\subsection{O$^0$ excitation and predicted line fluxes}

The excess flux in the \oishort\ line is limited in extent to the VLA\,1623 outflow, delineated by the rather uniformly distributed rotational \molh\ line emission. Especially at the LWS-position 150\asec\,NW are both \molh\ and [O\,I] emission directly seen to coincide spatially (Fig.\,\ref{h2_temp}). Inside the cloud, the atomic oxygen is neutral and the warm gas seen in the rotational \molh\ lines is expected therefore to contribute significantly to the [O\,I] emission detected by the LWS. 

Gas at temperatures around \powten{3}\,K can be expected to emit prolifically in oxygen fine structure lines  (see Fig.\,\ref{OI_flux}). Toward 150\asec\ NW, the observed flux ratio $F_{63}/F_{145} \gg 10$, which indicates that the emission is optically thin. However, we did in general not assume optically thin emission, but  treated the radiative transfer properly in the LVG approximation \citep[for details, see][]{liseau2006}.  We adopted the \molh-collisional rates from \citet{Jaquet1992} also for these computations and determined consistently the rates for the average ortho-to-para ratio of \molh\ and accounting for 20\% of neutral helium. 

For the analysis of the oxygen lines, we used the values of $T$ and $N$(\molh) determined from the observation of the \molh\ lines. Regarding the volume density, $n$(\molh), values in excess of $10^5$\,\cmthree\ are required by the very large flux ratio of the 63\,\um\ to 145\,\um\ lines, i.e. $n({\rm H}_2) \sim 3 \times 10^5$\,\cmthree\ (Fig.\,\ref{OI_ratio}). This value refers actually to the warm gas, but it is also identical to that determined by \citet{johnstone2000} from submm-observations of the cold dust. However, in the oxygen line analysis below, we examined a range in densities (and temperatures), whereas the oxygen abundance was kept constant at its ISM value, i.e. $3.4 \times 10^{-4}$ per H-nucleus \citep{oliveira2005}. 

\subsubsection{Cold gas phase oxygen}

A priori, one might not wish to dismiss the possibility that some or all of the [O\,I] line emission originates in the dense and cold gas, possibly associated with the high velocity CO gas, and we examine this case in the present section. 

As a conservative upper limit to the column density of the cold gas we take that for the corresponding average \av, i.e. $N({\rm H}_2) \le 3 \times 10^{22}$\,\cmtwo. This value is consistent with the findings also by others \citep[][and references therein]{larsson2007}. Assuming that the cloud depth is comparable to the width of the CO outflow, i.e. 50\asec\  or $9 \times 10^{16}$\,cm, the volume density is, again, $n({\rm H}_2)=3 \times 10^5$\,\cmthree.

The upper level energies for the \oishort\ and the \oilong\ lines are 228\,K and 326\,K above the ground. As also shown in Table\,\ref{OIlines}, oxygen gas at low temperatures, say $T=20$\,K, and with column density $N({\rm O}) \le 2 \times 10^{19}$\,\cmtwo\ can therefore not account for the observed emission, in particular not for that of the higher excitation \oilong\ line. As can be seen from this table, the 63\,\um\ line is very optically thick, $\tau_{63\,\mu {\rm m}}\sim 10^2$, which would result in highly uncertain estimates of the oxygen abundance. In contrast, the 145\,\um\ line is optically thin, $\tau_{145\,\mu {\rm m}}= 10^{-3}$, and the theoretical flux in that line can be expected to be more reliable. However, even for this large column density, the predicted line strength is inconsistent with the observation by large factors. Lowering the average density would increase these inconsistencies. In addition, for the densest part of the \roac, \citet{liseau2006} estimated an atomic oxygen column density  of only \powten{17}\,\cmtwo, decreasing the predicted line flux by another two orders of magnitude. In summary, cold gas emission, associated with that seen in low-lying rotational transitions of CO, does not contribute to the observed [O\,I] fluxes at any level of significance.

\subsubsection{Warm gas phase oxygen}

For the warm gas, we used the \molh\ parameters $T= 10^3$\,K and $N({\rm H}_2)=3.5 \times 10^{19}$\,\cmtwo, assuming an oxygen abundance of  $3.4 \times 10^{-4}$ relative to H. As a free parameter, the \molh\ density was varied in the range $10^{3}$ to \powten{6}\,\cmthree. We assumed the common linewidth of 1\,\kms\ for both lines (FWHM). This corresponds essentially to the thermal width at \powten{3}\,K, providing
maximum opacity in the lines. If associated with the outflow, their widths are likely broader. 

In Figs.\,\ref{OI_flux} and \ref{OI_ratio}, the dependencies of the [O\,I] line fluxes and their ratios on the temperature and density are shown. Displayed are the rather wide ranges of 300 - 3000\,K and $5 \times 10^{3}$ to \powten{6}\,\cmthree, respectively. Evidently, above the upper level energy of the \oilong\ line of 326\,K, the dependence of the line fluxes on the temperature becomes decreasingly weaker. For example, allowing for variations with, e.g., the standard deviation of 150\,K about the average temperature of \powten{3}\,K leads to changes in both lines by no more than 10\%. Therefore, temperature variations are unlikely to significantly alter the results of Table\,\ref{OIlines}. 

The line center optical depths are also presented in this table, demonstrating that the lines are indeed optically thin. Therefore, any signifcantly different value of the column density would directly alter the fluxes correspondingly. Such changes of $N$(\molh) would, however, be difficult to be reconciled with the \molh\ data.  

Also indicated in Fig.\,\ref{OI_flux} is the parameter space occupied by the observed \oishort\ line at the LWS position 150\asec\ NW and which corresponds to the $\pm 2 \sigma$ values of both the flux and the temperature. As it turns out, a unique solution is difficult to obtain and the results for two particular solutions are also presented in Table\,\ref{LWS_obs}. In the following, these will be examined in more detail.

\subsection{Extended low-density gas and standard O-abundance}

The first of these concerns gas at the rather low densities of below \powten{4}\,\cmthree\ and which is identified inside the dashed box of Fig.\,\ref{OI_flux}. This \oishort\ emission, filling the LWS beam, would also be in nice agreement with the observed \molh\ distribution. Sampled with 6\asec\ pixels, this \molh\ emission is clearly spatially resolved, showing some large-scale and coherent structure at \about\,\powten{3}\,K. In this scenario, the derived oxygen abundance would be essentially the one, which has been assumed by the calculations, viz. [O/H]\,$= 3.4 \times 10^{-4}$. Somewhat puzzling, though, is the non-detection of the accompanying 145\,\um\ line at the level of \powten{-12}\,\ecs\  (see Fig.\,\ref{OI_flux}). As a consequence, the observed flux ratio, $F_{63}/F_{145}$, is much larger than what can be accommodated by any low density gas, radiating at the observed flux level. 

Given the evidence provided by the other outflow positions observed with the LWS, it seems not justified to simply dismiss the validity of the 145\,\um\ data (see Table\,\ref{LWS_obs}). The S/N of these lines is not overwhelming, which could cast doubt on the goodness of the background corrected data. In general, when subtracting two nearly equal numbers, this creates a problem for the reliability of the result. However, in our particular case, systematically higher \oilong\ flux levels ($5.5 \times 10^{-12}$\,\ecs, rather than  $4.5 \times 10^{-12}$\,\ecs, prior to the background subtraction) should become easily detectable.     
 
\subsection{Compact high-density gas and standard O-abundance}

A second possibility concerns gas which actually fulfills the line-ratio requirement, i.e. gas which would have to be at densities exceeding \powten{5}\,\cmthree\ (Fig.\,\ref{OI_ratio}). The observation of similar ratios also at the other positions along the flow attests to the likely reality of this result.  Comparing to the data of Table\,\ref{OIlines}, a beam filling below the 10\% percent level would be indicated, which would correspond to linear source sizes of the order of 20\asec. This is not unreasonable, as small scale structure is observed in ro-vibrationally excited \molh\ emission \citep{dent1995,davis1999,caratti2006}, and also in [S\,II] lines from HH\,313 \citep{wilking1997,gomez2003,phelps2004}. 

These structures are small inhomogeneities (\lapprox\,10\asec) in the tenuous outflow gas, which are already pre-existent or have been broken off the cavity walls by Rayleigh-Taylor type instabilities. These experience impacts of essentially normal incidence and are shock-heated to several \powten{3}\,K. The non-detection of the [Ne\,II]\, 12.8\,\um\ line at 6\asec\ resolution  (Sect.\,2.1.2) limits the shock velocities to below 60\,\kms, however \citep{hollen1989}. These HH shocks have been modeled in detail by \citet{eisloffel2000}. Based on the observations in the NIR and in the optical, we know that their filling factors of the LWS beam are extremely small, i.e. \powten{-4} - \powten{-2}. We cannot exclude the possibility that these regions are larger, but hidden by extinction.  

For these compact sources of emission, derived temperatures range up to some thousand degrees (1\asec - 10\asec\ scale), higher than the \powten{3}\,K obtained in this paper (50\asec - 100\asec\ scale). However, theoretical fluxes would not change by much, as the temperature dependence remains essentially flat. For instance, the increase in [O\,I] line emission from high density gas at about 2000\,K would be below the 20\% level. The effects from a reasonable range of temperatures are well within the $2 \sigma$ flux boundaries considered in Fig.\,\ref{OI_flux} and would, as such, not significantly alter the results of Table\,\ref{OIlines}. 

\subsection{Extended high-density gas and low O-abundance}

A caveat of this latter scenario is the observed extent of the \molh\ emission as shown in Fig.\,\ref{h2_temp}. Line emission from accompanying atomic oxygen gas  would fill at least half of the LWS beam, i.e. more than what is indicated by the $F_{\rm obs}/F_{\rm pred}$\,\lapprox\,0.1 in Table\,\ref{OIlines}. This could potentially indicate a reduced oxygen abundance, i.e. formally to $2.4 \times 10^{-5}$, corresponding to a column density of $N({\rm O}) = 2 \times 10^{15}$\,\cmtwo. 

In this context, one issue concerns also the very origin of the atomic oxygen in this region. This gas is clearly associated with the CO outflow from VLA\,1623. This flow appears, at least partially, to be embedded in the \roca, the surface layers of which have been modeled as a PDR. According to the recent model by \citet{hollenbach2009} of the \ro\ PDR, the O$^0$ abundance is strongly reduced from its initial value at already quite modest values of \av\ into the cloud. These authors model successfully the Odin results for molecular oxygen, \molo, in \roa\ \citep{larsson2007}, the \molo\ being at its peak abundance. For their specific model and standard values\footnote{In \ro, the ortho-\water\ abundance (relative to the number of hydrogen nuclei) in the gas phase has variously been determined as $< 3.8 \times 10^{-7}$,  $2.2 \times 10^{-9}$  and $6.5 \times 10^{-7}$ \citep[][respectively]{liseauolofsson1999,ashby2000,franklin2008}. These values refer to an average over a large area and/or a large beam and it is conceivable that, along the outflow, \water\ abundances could locally be higher.} in their Eq.\,(32), we estimate an $X({\rm O}) = 2.4 \times 10^{-5}$. This particular value refers to a visual extinction \av\,\about\,4\,mag and for higher values, the oxygen abundance is predicted to decrease further. In addition, its coincidence with the value of the previous paragraph should be viewed as accidental, as the densities differ by a factor of six. 

To summarize, it is conceivable that the atomic oxygen observed in the outflow could be in dense cloud material with a reduced abundance compared to its initial ISM value. The emission would result from highly oblique shocks, as the flow is sliding along the cavity walls with shock velocities below 5\,\kms, resulting in temperatures of the dense wall gas up to \powten{3}\,K. Hardly any results from theoretical models for shock velocities below 5\,\kms\ are available \citep[e.g.,][]{kaufman1996}. The extended emission in pure-rotational \molh\ lines originates from this gas. This warm high density gas also dominates the [O\,I] emission.

\subsection{Summary: the oxygen abundance in \roa}

In conclusion, on the basis of the available observational evidence it is not possible to arrive at a unique solution to the oxygen abundance problem. If one disregards the \oilong\ data as too inaccurate, and therefore as irrelevant, the combined \molh\ and [O\,I] observations lead to an abundance which would be consistent with standard (depleted) ISM values. If, on the other hand, the \oilong\ data are invoked as relevant, the combined observations can be interpreted as indicating a considerably lower $X$(O). The latter scenario would be qualitatively in accord with recent models of cloud-PDR chemical evolution. To decide between these possibilities one would need to resolve the [O\,I] emission regions on scales smaller than 50\asec\ and, ideally, at the resolution of the \molh\ data with the ISO-CAM, i.e. 6\asec. At 63\,\um, the PACS instrument aboard the upcoming Herschel Space Observatory with its 9\asec\ pixels should come close enough and should be capable of providing this information.
 
\section{Conclusions}

Below, we briefly summarize our main conclusions, which are based on the interpretation of observations obtained with instruments aboard ISO.

\begin{itemize}
\item[$\bullet$] The analysis of ISOCAM-CVF spectra of the VLA\,1623 outflow revealed that \molh\ is appreciably excited mainly along the flow.
\item[$\bullet$] The emission in the six rotational lines S\,(2) to S\,(7) arises in gas of relatively constant temperature and column density.
\item[$\bullet$] Observations with the LWS along the flow and neighbouring regions show enhanced emission in the fine structure line \oishort\ in the warm outflow gas.
\item[$\bullet$] Invoking a model for the [O\,I] line excitation and radiative transfer leads to estimates of the atomic oxygen abundance in the dense core \roa. These calculations use primarily the parameters determined at 6\asec\ resolution for the \molh\ gas, but examine also a considerably wider range in temperature and density. 
\item[$\bullet$] This results in three main scenarios: (1) low density extended gas ($>70$\asec) with standard ISM oxygen abundance; (2)  high density compact sources ($< 20$\asec) with standard ISM oxygen abundance; (3) high density gas ($\ge 50$\asec) with largely reduced oxygen abundance.
\item[$\bullet$] The third option would be qualitatively in accord with recent models of chemical PDR-cloud evolution.
\item[$\bullet$] Observations with the Herschel Space Observatory can be expected to provide a test in the near future.  
\end{itemize}

\acknowledgement{We thank the referee for a stimulating discussion, the result of which certainly resulted in an improved manuscript. The contribution to this work in an initial phase by B.\,Larsson is acknowledged.}

\appendix

\section{The extinction curve}

A widely exploited extinction curve is the `average galactic law', calibrated for the near infrared by, e.g., \citet{rieke1985}. However, longward of about 5\,\um, no such generally accepted curve exists \citep[e.g.,][]{weingartner2001}. Dust properties may vary widely from location to location and unique extinction curves may be difficult to obtain. 

The optical depth of the extinguishing dust, at the wavelength $\lambda$ and along the line of sight, is given by 
$\tau_{\lambda} = 0.4 \,A_­{_{\lambda}}/\log{e}$, where the extinction at any wavelength, $A_­{_{\lambda}}$, is given in magnitudes. The corresponding extinction in the V-band, \av, can be found from $\tau_{\lambda} = 4.21 \times 10^{-5} \,\kappa_{\lambda} \,A_{\rm V}$, where 
$\kappa_{\lambda}$ is the mass extinction coefficient (absorption + scattering) in cm$^2$\,g$^{-1}$. For $\kappa_{\lambda}$ we adopt the values of \citet[][bare grains, MRN, $n$=0]{ossenkopf1994} and normalize the extinction in the K-band (2.2\,\um) through $A_{\rm K}=0.122$\,\av, assuming a gas-to-dust mass ratio of one hundred and the column density-\av\ relation of \citet{bohlin1978}, i.e., 
$N({\rm H\,I})+2 N({\rm H}_2) = 1.8\times 10^{21}\,A_{\rm V}\,\,({\rm in\,}{\rm cm}^{-2}\,{\rm mag}^{-1})$. Henceforth, we will refer to this extinction curve as the OHBG curve.

\begin{figure}[ht]
  \resizebox{\hsize}{!}{
  \rotatebox{00}{\includegraphics{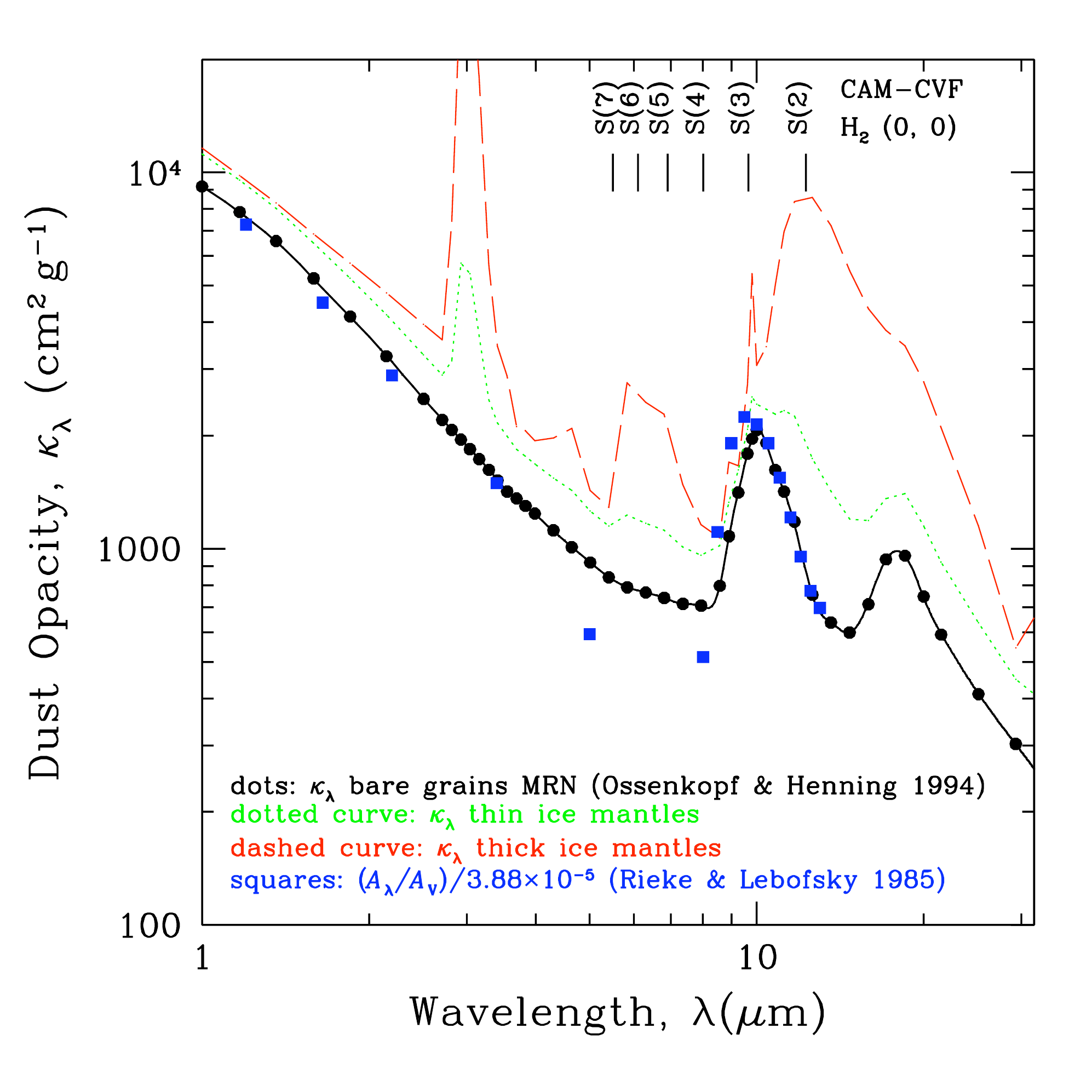}}
  }
  \caption{The extinction curve of \citet{ossenkopf1994} representing bare grains (OHBG) is shown as filled black dots and the interpolating solid line. The re-scaled data of \citet{rieke1985}, using $\kappa_{\lambda} = (A_{\lambda}/A_{\rm V})/3.88 \times 10^{-5}$, are shown as blue squares. Also shown are the models for ice-coated grains (green dots: thin ice mantles; red dashes: thick ice mantles). The wavelengths of the \molh\ lines admitted by the CAM-CVF are indicated by vertical bars. Toward the outflow from VLA\,1623, the observed absorption feature near 10\,\um\ has an average width of 1.4\,\um\ (FWHM), i.e., essentially that of the OHBG extinction curve. As is evident from the other curves, ice coatings would broaden that feature significantly.
  }
  \label{extinction}
\end{figure}

This coefficient for $A_{\rm K}$ for the OHBG curve is close to that given by \citet[][i.e., 0.112, see Fig.\,\ref{extinction}]{rieke1985}. It agrees excellently with the value suggested by \citet[][viz. 0.123]{savage1979} specifically for the \roc. The \citet{rieke1985}-curve is consistent with the $N({\rm H})-A_{\lambda}$ calibration in the I-band (0.9\,\um) suggested by \citet{cardelli1989} to be applicable also for dense clouds \citep[see also, ][]{kenyon1998} and provides, as such, a reasonable extrapolation into the infrared.

The extinction in the \roc\ is known to be `non-average', with a larger ratio of selective-to-total extinction \rv\,=\,\av/\ebv\,\gapprox\,5. The anomaly of the \ro\ extinction in the optical and the UV has commonly been explained as being the result of a preponderance of larger grains. This could be due to grain growth produced by coating of small grains with icy mantles. The low gas-phase \water\ abundance observed by SWAS \citep{ashby2000} would be in support for such a scenario and one might expect to see the growth effect by comparing observations with theoretical extinction curves for ice coated grains. However, it may be surprising to find that the observed width of the silicate absorption in \roa, i.e., FWHM(10\,\um)=1.4\,\um, is essentially that of `average' dust, i.e., significant broadening due to ice opacity is not observed (see Fig.\,\ref{extinction}). 

Whereas differences between individual extinction curves can become huge in the ultraviolet, errors introduced by a particular choice of curve are expected to become smaller in the infrared \citep{cardelli1989}.

For the OHBG curve, the extinction in units of magnitudes of \av\ was varied by steps of 0.1\,mag. All \molh\ lines in a given location (CAM-pixel) were assumed to have suffered the same amount of extinction. Linear regression fits to the \molh\ line flux data, with their error estimates, was made in a rotation diagram, resulting in a quantitative estimate of the goodness-of-fit, expressed equivalently by a $\chi^2$-value \citep{press1986}. The best fit was then selected as that with the smallest $\chi^2$, providing the corresponding \av\ for that pixel (see Fig.\,\ref{rot_diag}).

\section{\molh\ excitation}

\subsection{The single-temperature approximation}

The Herbig-Haro objects HH\,313\,A and B (Fig.\,\ref{VLA}) signal the presence of shocks with strong temperature gradients on relatively small spatial scales \citep{davis1999,eisloffel2000}. The extent of these localized regions is comparable to (or smaller than) the size of the 6\asec\ CAM-CVF pixel, i.e. \powten{16}\,cm. In a rotation diagram, the data would not follow straight lines but display significant curvature, i.e., upturns implied by the higher temperature gas. 

The fit of the \molh\ line data to a straight line would ignore any such curvature, which could be expected for a distribution in temperature within the pixel field of view. Significant amounts of gas at higher temperatures would result in significant flattening toward the higher upper-level energies and would result in lower quality fits (cf. Fig\,\ref{rot_diag}). This is not observed, however, when truncating the S\,(7) line from the data set with detections of all 6 \molh\ lines. The comparison of the 6-line and 5-line data sets in the region of overlap (63 pixels) demonstrates that the temperatures are identical within the observational errors. The overall mean temperature ratio is $1.00 \pm 0.07$.

Selecting the data set with only the S\,(2) to S\,(6) detections nearly triples the number of pixels from 63 to 162, significantly enlarging the region of \molh\ emission which can be analysed, i.e. $\sqrt{162} \times 6$\asec, which is comparable to the angle subtended by the LWS \citep[\about 70\asec,][]{gry2003}. Remarkably, also over this extended scale, the gas temperatures exhibit a uniform distribution: whereas, e.g., the extinction varies widely between a few and several tens of magnitudes of \av, the temperatures stay rather flat at \about\,\powten{3}\,K (Fig.\,\ref{T_AV_Sil}). Hence, we conclude that the extended warm \molh\ outflow gas near VLA\,1623 is reasonably well represented by the single temperature approximation (see also Fig.\,\ref{h2_temp}).

\subsection{Thermal equilibrium of the \molh-molecules}

\begin{figure}[ht]
  \resizebox{\hsize}{!}{
  \rotatebox{00}{\includegraphics{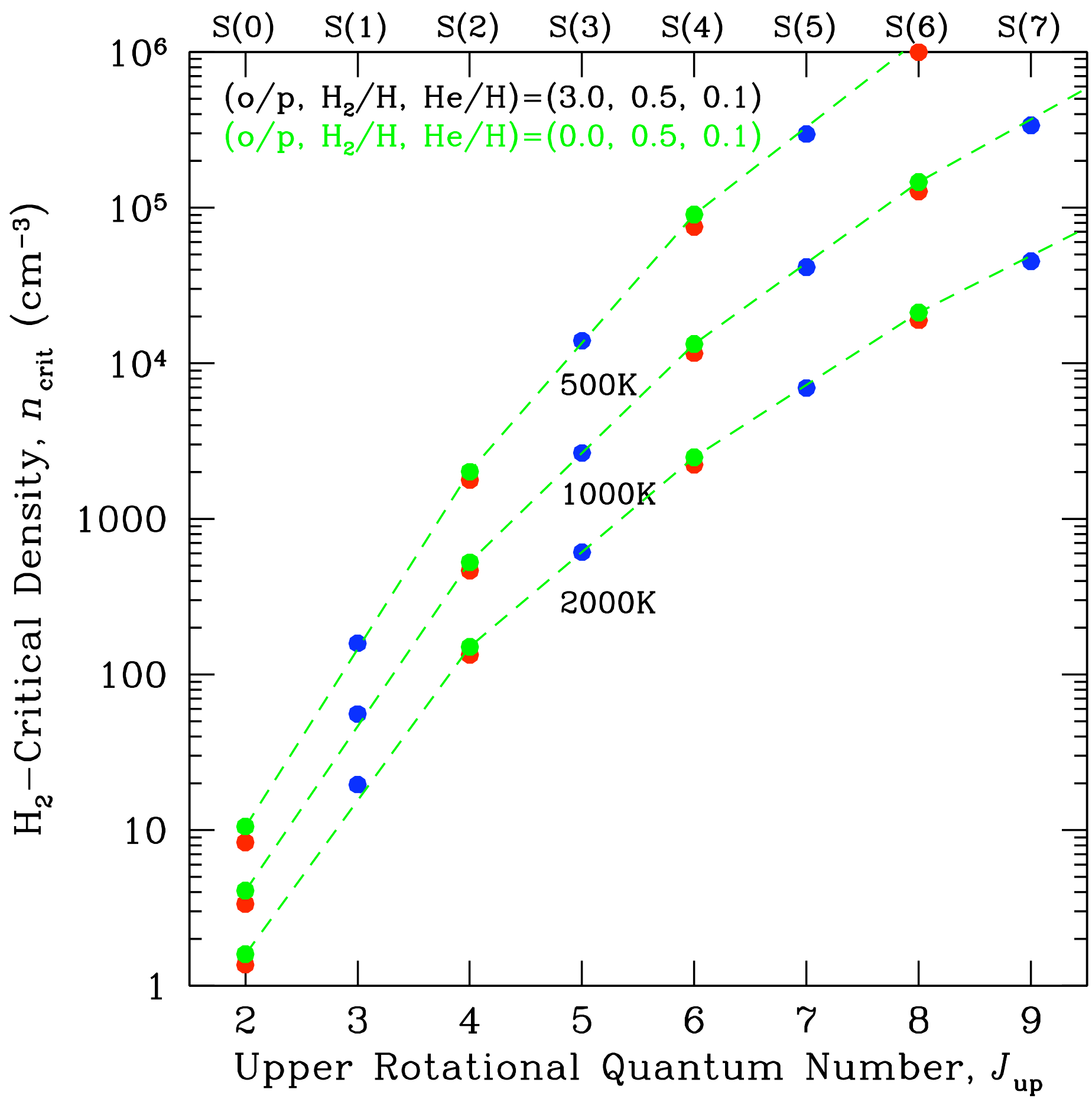}}
  }
  \caption{The critical density, $n_{\rm crit}=A_{\rm ul}/\gamma_{\rm ul}(T)$, for rotational transitions in the vibrational ground state, (\vbis, \vprim\,= 0, 0), for three values of the temperature, $T$. Considered are collisions with H\,I, He, ortho-\molh\ and para-\molh. The two extreme cases, i.e., for o/p = 3 (red and blue symbols) and o/p = 0 (green dashes), respectively, are shown. Connecting lines have no physical meaning and are inserted for clarity only. In addition to the ortho-to-para ratio, parameters are $X$(\molh) = 0.5 (all hydrogen molecular) and $X$(He) = 0.1. 
  }
  \label{ncrit}
\end{figure}

The level populations become thermalized at densities which are higher than the `critical density', $n_{\rm crit}=A_{\rm ul}/\gamma_{\rm ul}(T)$, where radiative and collisional de-excitation rates are equal (Fig.\,\ref{ncrit}). In that figure, Einstein-$A$ values have been adopted from \citet{wolniewicz1998} and the rate coefficients for collisional de-excitation, $\gamma_{\rm ul}(T)$, were taken from the web-site maintained by D.\,Flower, \texttt{http://ccp7.dur.ac.uk/ccp7/cooling\_by\_h2/} \citep[see][]{lebourlot1999}. Data are available for collisions with H\,I, He, ortho-\molh\ and para-\molh. The total collisional rate at a given temperature is therefore the sum of the fractional contributions by the collision partners.

\roa\ is a {\it dense core}, so that densities can be expected to be high and that the assumption of LTE therefore would be plausible, at least for the lower $J$-transitions (Fig.\,\ref{ncrit}). Generally, in regions, where temperatures would be relatively low ($T \sim 500$\,K), the critical density of the S\,(7) line would exceed \powten{6}\,\cmthree\ and the line would likely not be thermalized. In contrast, at the nominal temperature of \powten{3}\,K in \roa,  $n_{\rm crit}$ for the S\,(7) line would be below the average density of the VLA\,1623 region, i.e., $n({\rm H}_2)= 3 \times 10^5$\,\cmthree\ (see Sects.\,3.5 \& 3.6) and we conclude that the assumption of LTE is justified.

\subsection{The optical depth of the \molh-lines}

It is easy to show that the spectral lines are optically thin, $\tau(\upsilon_0) \ll 1$, as long as the column density of warm \molh\ gas obeys $N({\rm H}_2) \ll 1.5 \times 10^{22}\,\,{\rm cm}^{-2}$. 

The numerical coefficient is based on the following: $A_{\rm ul} \propto  \nu_0^3$, $\max({n_{\rm u}/n_{\rm tot}})=1.0$, $\min{\Delta \upsilon}=1.0$\,\kms, $\min{T}=h \nu_0/k \approx 500$\,K for the S(0) line and $\max{X({\rm H}_2)} \equiv \max{N({\rm H}_2)/N({\rm H)}}=0.5$. Realistic values for higher transitions will always result in lower column densities, hence smaller optical depths ($T\ge 500$\,K) and, for the proper parameters, this condition is always checked a posteriori. Also, in outflow gas, $\Delta \upsilon \gg 1$\,\kms, decreasing $\tau(\upsilon_0)$ per unit column even further. The condition of optically thin emission is clearly fulfilled.

\end{document}